# Split-ball resonator

*Arseniy I. Kuznetsov,[1,*] Andrey E. Miroshnichenko,[2] Yuan Hsing Fu,[1] Vignesh Viswanathan,[3] Mohsen Rahmani,[1] Vytautas Valuckas,[1,3] Zhen Ying Pan,[1] Yuri Kivshar,[2] Daniel S. Pickard,[3] and Boris Luk'yanchuk,[1]*

1. Data Storage Institute, A*STAR (Agency for Science, Technology and Research), 5 Engineering Drive 1, 117608, Singapore

2. Nonlinear Physics Centre, Research School of Physics and Engineering, Australian National University, Canberra, ACT 0200, Australia;

3. Department of Electrical and Computer Engineering, National University of Singapore, 1 Engineering Drive 2, 117576, Singapore

*Corresponding author: arseniy_k@dsi.a-star.edu.sg

**Abstract:** We introduce a new concept of *split-ball resonator* and demonstrate a strong magnetic dipole response for both gold and silver spherical plasmonic nanoparticles with nanometer-scale cuts. Tunability of the magnetic dipole resonance throughout the visible spectral range is demonstrated by changing the depth and width of the nanoscale cut. We realize this novel concept experimentally by employing the laser-induced transfer method to produce near-perfect spheres and helium ion beam milling to make cuts with the nanometer resolution. Due to high quality of the spherical particle shape governed by strong surface tension forces during the laser transfer process and clean, straight side walls of the cut made by helium ion milling, the magnetic resonance is observed at 600 nm in gold and at 565 nm in silver nanoparticles. Structuring arbitrary features on the surface of nanoscale spherical resonators provides new ways for engineering hybrid resonant modes and ultra-high near-field enhancement.

One of the main challenges in the field of plasmonics and metamaterials during the last decade is to engineer nanostructures with strong magnetic and electric dipole resonances at optical frequencies [1-5]. Getting these two resonances together in the same frequency range can lead to unique material properties associated with near-zero or even negative effective refractive index. The concept of split-ring resonator (SRR), which provides strong magnetic dipole response of metallic structures, was theoretically introduces by Pendry et al. in 1999 [6]. Since then many efforts to experimentally demonstrate magnetic resonance of metallic structures has been reported, first at GHz [7], then at THz [8], and finally at optical frequencies [9, 10]. It has been shown that scaling down the sizes of the split-ring resonator linearly increases the magnetic resonance frequency. However, this linear dependence saturates close to the visible spectral range mainly due to non-ideality of plasmonic metals at these frequencies [10, 11]. Further control of the resonance position is possible using optimization of the design of SRRs [12, 13]. More advanced designs, such as cut-wire pairs [14, 15], fish-net structures [16, 17], and nanoparticle clusters [18, 19], have also been implemented to facilitate metamaterial fabrication and further increase their magnetic resonance frequency. However, until now there were only several experimental demonstrations of resonant magnetic response of metallic nanostructures in the visible spectrum [18-25]. Strong losses of plasmonic structures in the visible spectral range significantly limit the possibilities of shifting the magnetic resonance to higher frequencies. These losses arise due to imperfections of the nanostructures at sub-100nm scale and intrinsic losses inherent to plasmonic metals in this frequency range. The use of low-loss resonant high-refractive index dielectric nanoparticles having strong Mie resonances of both electric and magnetic nature [26-32] can partially solve this problem and shift the magnetic resonance frequency up to the blue spectral range [26, 28, 31]. However, lower field enhancement around such nanoparticles compared to metals makes them only a partial substitution for plasmonics.

In this paper, we introduce and experimentally demonstrate a new concept of a strong magnetic dipole resonance tuneable almost throughout the whole visible spectral range using standard plasmonic metals such as gold and silver. The key aspect is a nanometer-size cut fabricated inside an almost perfectly spherical plasmonic nanoparticle. Such 3D spherical design allows shifting magnetic dipole resonance down to electric dipole resonance wavelength in the visible spectral range. Experimentally this novel design is realized by combining the almost perfect spherical nanoparticles achievable with laser-induced transfer method (LIT) with the unprecedented patterning fidelity at nanometer-scale dimensions characteristic of milling with a focused helium ion beam. Near-field enhancement and Poynting vector flow inside these unique nanostructures are also discussed.

**Results**

*Numerical simulations*

Schematic representation of a metallic sphere with a cut, split-ball resonator (SBR), is shown in Fig.1. The nanocut is introduced on top of the sphere making it similar to 3D upright split-ring resonators [33]. In this case, the coupling of light to the magnetic resonance mode should be more efficient due to both electric and magnetic fields contribution.

To analyze abilities of the SBR to support magnetic dipole resonance we performed numerical simulations of its optical spectral properties using commercial CST Microwave Studio software. Results are shown in Fig.2 for a gold sphere with diameter of 200 nm having a 50 nm wide cut going down to the center of the sphere. The nanostructure is located in free space and excited by a plane wave directed from the top (antiparallel to Z axis) with two different polarizations, parallel (along Y axis) and perpendicular (along X axis) to the nanocut direction. Total scattering and absorption efficiencies by the nanostructure are shown in Fig. 2a (see section Methods for details).

When the incident light polarization is parallel to the cut, only a single electric dipole resonance is excited inside the SBR, similar to a spherical nanoparticle without a cut (Fig.2a). The electric dipole nature of this resonant peak is confirmed by electric and magnetic field distribution at the resonance wavelength (Fig.2b&c) and decomposition of the near-field into spherical multipoles (Fig.2g, see detailed discussions below). When incident light polarization is perpendicular to the cut, two resonant peaks appear in the scattering spectrum: the first weaker peak is close to the position of the electric resonance, and the second, much stronger peak is at the longer wavelengths (see Fig. 2a). This second resonant peak corresponds to the so-called *LC resonance* excited in SBR similar to the case of split-ring resonators [6-13, 34]. The scattering intensity at this resonance is defined by a combination of strong electric and magnetic dipole contributions (Fig. 2f, see detailed discussions below). Electric field at this resonant wavelength is strongly enhanced inside the cut generating circulating electric currents flowing around the nanocut inside the nanoparticle (Fig.2d). On the other hand the magnetic field is strongly enhanced and oscillates inside the nanocut parallel to its axis (Fig.2e).

Similar spectral behavior can also be observed in the SBR excited from the side facing the nanocut (along Y axis) with light polarization parallel (along Z axis) and perpendicular (along X axis) to the cut (Fig.2a, dashed lines). Intensities and spectral positions of both electric and magnetic *LC* resonances in this case are almost unchanged compared to the case of excitation from the top. This shows that the magnetic response of SBRs is almost independent on the direction of incidence when it is varied in the nanocut plane and the incident light polarization is perpendicular to the gap. This property is similar to conventional split-ring resonators, and it can be advantageous compared to multi-layer fishnets and stripes for a design of isotropic metamaterials.

To understand better the nature of resonances excited inside SBR for both polarizations, we perform a decomposition of the scattered field of the nanostructure into multipole moments based on the vector spherical harmonics [34-36] (see section Methods for details). Results of this analysis for different incident angles and light polarizations are presented in Fig. 2(f)-(i). Notation $b_{10}$ corresponds to the magnetic dipole mode, while $a_{10}$ and $a_{11}$ correspond to the electric dipole modes of two different types. For the incident light polarization parallel to the nanocut, only a single electric resonant mode is excited while the excitation of other modes is negligibly small (Fig.2g&i). This conclusion fully supports our previous discussion based on the electric and magnetic field distributions inside SBR (Fig. 2b&c). We note here that the type of the electric mode changes for different polarizations from $a_{10}$ for incidence from the top (Fig.2g) to $a_{11}$, for incidence from the side (Fig.2i) of SBR. For polarization perpendicular to the nanocut, the resonance around 900 nm exhibits an enhancement of the magnetic dipole response $b_{10}$ for both incident angles (see Fig. 2f&h). However, the electric dipole mode $a_{11}$ is also resonantly excited at this wavelength, which is caused by a strong electric field enhancement inside the nanocut (see Fig.2d). This situation differs strongly from resonant dielectric nanoparticles where a pure magnetic dipole mode can be excited [26-32]. However, it is similar to conventional split-ring resonators, where electric field enhancement inside the gap contributes into the scattering [34]. Relative contribution of the magnetic dipole mode to the scattering efficiency at *LC* resonance with respect to the electric dipole mode in SBR is comparable and even stronger with that for conventional SRRs [34] (see Supplementary Fig. S1).

In Supplementary Information we also present the analysis of similar modes for light scattering by two split-ring resonator structures with the same size of the nanogap. One of the structures is a nanocube with the side length of 200nm, gap width of 50nm and depth of 100nm, which is a 3D nanostructure similar to the SBR. The second is a flat split-ring resonator obtained from the nanocube by shrinking its width down to 20 nm. Both of the nanostructures exhibit *LC*-type resonance at longer wavelengths, but with smaller magnetic mode contribution compared to SBR. This allows us to claim that the split-ball resonator supports strong magnetic dipole mode at *LC*-type resonance. It is also important to note that relative contribution of magnetic dipole to the *LC* resonance is somewhat higher in 3D cube-like split-ring structure than in the flat SRR. This shows advantage of 3D designs over flat nanostructures for generation of strong magnetic dipole response.

To show how both electric and magnetic resonances behave with respect to the cut parameters of the SBR, different cut depths have been simulated at a fixed cut width of 20 nm (Fig.3a). The cut width has also been gradually changed at a fixed cut depth of 100 nm (Fig.3b). As it is seen from the figures both parameters strongly affects the *LC* resonance spectral position, while the position of the short-wavelength electric resonance is almost unchanged though its intensity is changing. When the cut depth is decreased the *LC* resonance is blue-shifted, reaching the electric resonance position in the visible spectral range (Fig.3a). This can be explained by decrease of the capacitance of the gap (*C*) leading to increase of the resonance frequency ($\omega^2 = 1/LC$) which is in line with the split-ring resonators theory [9-11]. On the other hand decrease of the width of the gap increases the gap capacitance, which leads to decrease of the resonance frequency and a corresponding redshift of the *LC* resonance peak (Fig. 3b). In both these cases the inductance (*L*) changes slowly due to the large size of the sphere compared to the gap size. One should also note that the above *LC*-circuit model initially developed for large split-ring resonators is not fully applicable to our case of very small nanoparticles with nanocuts and is only used here as a simple illustration of the observed effects. Similar behavior with respect to the depth of the gap has earlier been observed in flat split ring resonators fabricated by focused ion beam milling of thin gold films [12]. It was shown that controlling the depth of the gap it is possible to shift the *LC* resonance down to the electric resonance wavelength. However, as the structures were flat the position of both resonances was in the near-IR spectral range. In our case, the electric resonance position of the spherical particles is located in the visible spectrum which gives a possibility to shift the magnetic *LC* resonance down to visible wavelengths.

*Experiment*

To verify this concept experimentally we first fabricated gold and silver nanoparticles using laser-induced transfer method in combination with e-beam lithography [37] (see section Methods for more details). This method allows producing nanoparticles with almost perfect spherical shape and high surface quality. The size of the nanoparticles is controlled using initial lithography step while their spherical shape and low surface roughness is assured by strong surface tension forces of molten metal during the laser processing. Then nanocuts with straight side walls have been produced on top of the nanoparticles using focused helium (He) ion beam milling. In comparison to standard FIB systems [38, 39], which use gallium (Ga) ions, HIM can provide significantly higher structuring resolution [40-42]. Also, in contrast to standard gallium-ions based FIB systems, HIM does not significantly dope side walls of the materials during milling and thus may keep good plasmonic properties of the nanoparticles [38, 43].

HIM images of two gold nanoparticles with cuts fabricated by this combined method are shown in Fig.4 a&b. These nanoparticles have a similar diameter of around 170 nm and cut width of around 15 nm. The structured cut can be controlled in depth and width (down to <5nm) but in this case was varied to match the modeling predictions. Variable depth was achieved by varying the dwell time at a fixed beam current. The dwell time was 25% longer for the particle shown in Fig.4a compared to that shown in Fig.4b which resulted in a deeper cut (see section Methods for details). Optical properties of these nanoparticles have been studied using single-nanoparticle spectroscopy setup [28]. Measurements were performed in dark-field geometry with illumination from one side facing the nanocut (*k* vector is in YZ plane). Two different polarizations of excitation source have been studied: parallel and perpendicular to the nanocut (see Fig.1 and section Methods for details). In this excitation geometry, p-polarization (lying in YZ plane) should only couple to electric dipole resonance of the sphere while s-polarization (along X axis) should excite both electric and magnetic dipoles. Results of these measurements for the two gold particles are shown in Fig.4 c&d. For p-polarization, which is parallel to the cut, only a single scattering peak corresponding to the electric dipole resonance of the sphere is observed. For s-polarization, two resonances are detected, *LC* resonance on the long-wavelength side and a left-over of electric dipole on the short-wavelength side. For the particle with a deeper cut (Fig.4a&c) the *LC* resonance position is red shifted compared to the particle with a shallower cut (Fig.4b&d). These results are in a very good agreement with theoretical predictions discussed in the previous section (Fig.3a). To the best of our knowledge, the wavelength of magnetic resonance of 600 nm obtained for the gold SBR in Fig.4b&d is the shortest compared to all gold-based nanostructures reported so far.

To show that this concept may also work at even shorter wavelengths the same approach has been applied to fabricate SBRs made of silver. Plasmonic resonance of silver nanoparticles is blue-shifted compared to gold. Thus one may expect further blue shift of the magnetic resonance position in silver SBRs. Results of this study are shown in Fig.5. First, a silver SBR with a particle diameter of around 170 nm and a cut width of around 15 nm, which is similar to the earlier discussed gold SBRs, has been fabricated (Fig.5a). The *LC* resonance position in this case is around 640 nm (Fig.5d), which is similar to the above results for gold nanoparticles. However, this resonance is more pronounced and well separated from the electric dipole resonance located around 500 nm. To further shift both resonances to shorter wavelengths a smaller silver nanoparticle with diameter of around 80 nm and a shallow cut of around 12 nm width has been fabricated (Fig.5b). In this case, the *LC* resonance position is further blue-shifted reaching the wavelength of 570 nm, which is noticeably smaller than earlier reported results obtained with split-ring resonators [10-13] and coupled-nanoparticles [18, 19] and is comparable to the best results obtained with multi-layer stripes and fishnets [21, 22]. However, in contrast to the stripes, fish-nets and other multi-layer structures whose magnetic response can only be obtained under a single excitation direction, split-ball resonators (similar to SRRs) possess an omnidirectional magnetic response in the incidence plane parallel to the nanocut, which may hold a bigger promise for metamaterial design and fabrication.

To show how the surface quality of the nanoparticles influences their resonant properties, HIM cut have also been introduced into a commercial chemically synthesized silver nanoparticle (*NanoComposix*) of 80 nm size (Fig.5c&f). In this case, the electric resonance position is further blue-shifted, which may probably be explained by a higher crystallinity of the chemically synthetized silver nanoparticles. However, the *LC* resonance is less pronounced due to the inferior surface quality of these nanoparticles compared to the nanoparticles obtained by LIT method.

**Discussion**

Finally, light energy flow inside and outside the SBR at the magnetic resonance wavelength has been simulated by CST Microwave Studio software. This study is useful to understand the high scattering and absorption efficiencies of the nanostructures. Similar SBR parameters as in Fig.2 have been taken for simulations. The Poynting vector lines at the *LC* resonance wavelength of 882 nm are shown in Fig.6a and b in two different cross-section planes. The color map corresponds to the distribution of the modulus of the Pointing vector. It is seen from Fig.6a that in the XZ plane the Poynting vector lines spanning an area exceeding the nanoparticle cross-section by several times are collected and concentrated inside the nanocut. This generates significant field enhancement of the order of 20 inside the cut. Similar simulations performed for a silver SBR with diameter of 170 nm and a cut width of 15 nm (similar to the one shown in Fig.5a) demonstrate even larger field enhancement inside the cut around 45 (not shown here). This energy flow passing the cut does not come out of the SBR forming optical vortices from both sides of the nanostructure. This explains strong absorption resonance at this wavelength. On the other hand, in YZ plane most of the Poynting vector lines are strongly deviated and passing around the SBR without penetrating into it, resulting in a strong scattering efficiency at the *LC* resonance. Fig.2c shows separatrix absorption surface of the SBR in XY plane. All Poynting vector lines inside the separatrix surface go into the SBR and are absorbed inside [44, 45]. Strong asymmetry of the separatrix surface in X and Y direction supports our discussion above.

As a comparison, energy flow at the wavelength of the minimum scattering of 785 nm is also shown in XZ and YZ planes in Fig.6d and e respectively. In this case, the Poynting vector lines pass through the SBR with only a very small absorption and scattering making the nanostructure almost invisible. The separatrix surface in this case (Fig.6f) has a different geometry compared to the resonance (Fig.6c) and is smaller than the geometrical cross-section of the particle. This behavior is associated with destructive Fano-type interference between the electric dipole and *LC* resonance inside SBR [18, 46-48].

Strong field-enhancement and magnetic dipole resonance of the split-ball resonators make them promising for future applications in metamaterials, surface-enhanced Raman scattering (SERS), heat-assisted magnetic recording (HAMR), and nanoantennas.

In conclusion, a novel concept to realize optical magnetism at visible frequencies with standard plasmonic materials such as gold and silver is proposed. This concept is based on a spherical nanoparticle with a precision patterned nanocut. Due to a volumetric 3D shape of the proposed design a strong magnetic response tunable through the visible spectral range can be obtained. Experimentally this unique design was realized using laser-induced melting of nanostructures to obtain an almost perfect spherical nanoparticles and helium ion beam milling to make cuts with nanometer resolution and straight side walls. Magnetic resonance wavelength down to 600 nm for gold and 565 nm for silver is experimentally demonstrated, which corresponds and exceeds previous experimental results. This novel approach to engineer resonant modes of perfect spherical nanoparticles at nanoscale opens new possibilities to design novel nanostructures with unique modes and high field enhancement regions which may be promising for applications to metamaterials, surface-enhanced Raman scattering (SERS), heat-assisted magnetic recording (HAMR), and nanoantennas.

**Methods**

*E-beam nanofabrication*

Gold and silver nanodisks with various diameters were fabricated on the quartz substrates by electron beam lithography (Elonix 100KV EBL system). First 120 nm positive resist (ZEP) was coated on the sample. After baking it at 180 °C for 2 min, a 20 nm Espacer was coated on the top of the resist to avoid charging effect. The structural geometries were defined by electron-beam in the resist followed by a standard development. Then a thin Ti film (2 nm thick) was deposited on the substrate by e-beam evaporation to ensure good adhesion between the evaporated in the following 50 nm Au/Ag layers and the substrate. Lift-off procedure was the last step to generate the nanodisks.

*Laser processing*

Following the first nanofabrication steps femtosecond laser irradiation has been applied to the nanostructures. In these experiments, commercial 1 kHz femtosecond laser system (Tsunami+Spitfire, Spectra Physics) delivering 1 mJ, 100 fs laser pulses at a central wavelength of 800 nm has been used. The laser beam with diameter of 4 mm was directed onto a square-shaped pinhole with 300 μm size. The beam just after passing through the pinhole was imaged onto the sample surface using a 20× long-distance microscope objective (Mitutoyo, MPlan NIR 20) and a 200 mm focus tube lens. This image transfer results in a square-shape 15 μm size flat-top laser beam profile on the sample surface. The laser beam power was attenuated by a neutral-glass filter (15%) and a half-wave plate + polarizer to get the fluence of around 0.5 J/cm$^2$ on the sample surface. The nanostructures have been melted by the laser irradiation, transformed into spherical nanodroplets by surface tension forces and transferred onto an ITO-coated glass substrate. After solidification the molten droplets form almost perfect spherical nanoparticles. More details about laser-induced transfer method and its application to nanostructures and continuous metal films for nanoparticle generation can be found in [37, 49-51].

*Helium ion beam milling*

The Helium Ion Microscope (HIM) (Zeiss Orion Plus) was employed in conjunction with a Nabity Pattern Generator to further structure the surface of the fabricated nanoparticles. HIM is capable of milling extremely high resolution structures with lateral dimensions of less than 5nm. When combined with a sophisticated pattern generator and custom writing strategies, near arbitrary control of the surface features is possible. In this work, the ion fluence across the cut was varied and optimized with a multi-pass writing strategy to prevent over-milling at the edges, and eliminate residual metal redeposition while maintaining the pattern fidelity. The beam dwell time was modulated throughout the length of the cut, with commensurately longer milling time at the center of the cut compared to the edges to compensate for the center's increased metal thickness, as well as mitigate secondary effects such as decreased redeposition at the edges. For gold particle in Fig.4a, the ion fluence was about $5\times10^{16}$ ions/cm$^2$ at the edges with intermediate sections having a varying fluence of $7.5\times10^{17}$, $3.25\times10^{18}$, $5.25\times10^{18}$ and $6.75\times10^{18}$ ions/cm$^2$ towards the center. The ion fluence was decreased by 25% for particle in Fig.4b. For the silver particle in Fig.5a, the fluence at the edges was about $8\times10^{17}$ ions/cm$^2$ with intermediate sections having a range of $2\times10^{18}$ to $5\times10^{18}$ ions/cm$^2$ and $8\times10^{18}$ ions/cm2 at the center. Silver particles in Fig.5b&c were patterned with fluence scaled down by 3 times. The helium ion beam landing energy of ~ 35keV with ion beam current of ~0.5 pA was employed.

*Single nanoparticle spectroscopy*

Spectral analysis of the fabricated split-ball resonators was performed using single-nanoparticle spectroscopy setup (see [28] for details) in dark-field geometry (Fig.1). The sample was irradiated by a halogen lamp source from one side facing the gap at an angle of 58.5° to the surface normal (see [31] for details of dark-field spectral measurements). The scattering by the nanostructure has been collected from top into a solid angle corresponding to the microscope objective with 0.55 NA. Two different incident light polarizations have been studied: s-polarization – perpendicular to the nanocut and parallel to the substrate surface, and p-polarization – parallel to the nanocut and at an angle to the substrate surface (Fig.1). The collected scattering spectra for both polarizations are normalized to the halogen lamp spectrum measured in bright-field transmission geometry. This normalization method provides correct shapes of the scattering spectra without keeping information on total scattering amplitudes.

*Theoretical and numerical analysis*

Numerical simulations of the SBRs were performed using commercial CST Microwave Studio software. Scattering and absorption efficiencies by the nanostructures were obtained by dividing the corresponding cross-sections by overall geometrical cross-section of the nanoparticle.

To calculate relative contributions of different modes excited inside SBR, we perform a decomposition of the scattered field of the nanostructure into multipole moments based on the vector spherical harmonics [34-36]. To do that we calculated electric $a_{lm}$ and magnetic $b_{lm}$ multiople coefficients via the electric scattered field $\mathbf{E}_s$ on a spherical surface enclosing the split ball centered at the origin (center of the split ball) as follows:

$$a_{lm} = \frac{(-i)^{l+1} kR}{h_l^{(1)}(kR) E_0 \sqrt{\pi(2l+1)l(l+1)}} \int_0^{2\pi}\int_0^{\pi} Y_{lm}^*(\theta,\phi)\hat{\mathbf{r}} \cdot \mathbf{E}_s(\mathbf{r}) \sin\theta\, d\theta\, d\phi$$

$$b_{lm} = \frac{(-i)^{l} kR}{h_l^{(1)}(kR) E_0 \sqrt{\pi(2l+1)}} \int_0^{2\pi}\int_0^{\pi} \mathbf{X}_{lm}^*(\theta,\phi) \cdot \mathbf{E}_s(\mathbf{r}) \sin\theta\, d\theta\, d\phi$$

where $R$ is the radius of the enclosing sphere, $k$ is the wavevector, $h_l^{(1)}$ is the Hankel function with the asymptotics of the outgoing spherical wave, $E_0$ is the amplitude of the incident wave, $Y_{lm}$ and $\mathbf{X}_{lm}$ are scalar and vector spherical harmonics [36]. Based on these multipole coefficients we can also calculate partial scattering efficiencies associated with a particular spherical harmonic

$$Q_{lm}^{sca,E} = \frac{1}{(kd)^2}(2l+1)|a_{lm}|^2 \quad , \quad Q_{lm}^{sca,H} = \frac{1}{(kd)^2}(2l+1)|b_{lm}|^2,$$

where $d$ is the outer radius of the split ball. The total scattering efficiency can be obtained by taking the full sum of all partial contributions:

$$Q^{sca} = \sum_{l=1}^{\infty} \sum_{m=-l}^{l} Q_{lm}^{sca,E} + Q_{lm}^{sca,H}$$

Although our system does not exhibit pure azimuthal symmetry, the numerical results demonstrate that mulitpole coefficients still hold the following relations: $|a_{l,m}| = |a_{l,-m}|$ and $|b_{l,m}| = |b_{l,-m}|$ which can be explained due to relatively narrow nanocut of the split ball. Based on these results we can introduce the effective scattering efficiencies

$$\tilde{Q}_{lm}^{sca,E} = Q_{l,m}^{sca,E} + Q_{l,-m}^{sca,E} = \frac{2}{(kd)^2}(2l+1)|a_{lm}|^2 \quad , \quad \tilde{Q}_{lm}^{sca,H} = Q_{l,m}^{sca,H} + Q_{l,-m}^{sca,H} = \frac{2}{(kd)^2}(2l+1)|b_{lm}|^2,$$

which are the sum of two contribution for ($l,m$) and ($l,–m$) harmonics for $m \neq 0$. These efficiencies were plotted in Fig. 2(f)-(i) for all electric and magnetic modes of the first and the second order (dipoles and quadrupoles). Electric and magnetic dipole modes, which have the strongest contribution to the total scattering, are marked in Fig.2 as: $b_{10}$ – magnetic dipole mode, $a_{10}$ and $a_{11}$ – electric dipole modes.

**References**

1. Soukoulis, C. M & Wegener, M. Past achievements and future challenges in the development of three-dimensional photonic metamaterials. *Nature Photon.* **5**, 523-530 (2011).

2. Boltasseva, A. & Atwater, H. A. Low-loss plasmonic metamaterials. *Science* **331,** 290-291 (2011).

3. Soukoulis, C. M. & Wegener, M. Optical metamaterials – more bulky and less lossy. *Science* **330,** 1633-1634 (2010).

4. Zheludev, N. I. The road ahead for metamaterials. *Science* **328,** 582-583 (2010).

5. Shalaev, V. M. Optical negative-index metamaterials. *Nature Photon.* **1,** 41-47 (2007).

6. Pendry, J. B. , Holden, A. J., Robbins, D. J. & Stewart, W. J. Magnetism from conductors and enhanced nonlinear phenomena. *IEEE Trans. Microwave Theory Tech.* **47,** 2075-2084 (1999).

7. Smith, D. R., Padilla, W. J., Vier, D. C., Nemat-Nasser, S. C. & Schultz, S. Composite medium with simultaneously negative permeability and permittivity. *Phys. Rev. Lett.* **84,** 4184- 4187 (2000).

8. Yen, T. J., Padilla, W. J., Fang, N., Vier, D. C., Smith, D. R., Pendry, J. B., Basov, D. N. & Zhang, X. Terahertz magnetic response from artificial materials. *Science* **303,** 1494-1496 (2004).

9. Linden, S., Enkrich, C., Dolling, G., Klein, M. W., Zhou, J., Koschny, T., Soukoulis, C. M., Burger, S., Schmidt F. & Wegener, M. Photonic metamaterials: magnetism at optical frequencies. *IEEE J. Sel. Top. Quant. Electron.* **12**, 1097-1105 (2006).

10. Klein, M. W., Enkrich, C., Wegener, M., Soukoulis, C. M. & Linden, S. Single-slit split-ring resonators at optical frequencies: limits of size scaling. *Opt. Lett.* **31**, 1259 (2006).

11. Zhou, J., Koschny, T., Kafesaki, M., Economou, E. N., Pendry, J. B. & Soukoulis, C. M. Saturation of the magnetic response of split-ring resonators at optical frequencies. *Phys. Rev. Lett.* **95**, 223902 (2005).

12. Enkrich, C., Perez-Willard, F., Gerthsen, D., Zhou, J., Koschny, T., Soukoulis, C. M., Wegener, M. & Linden, S. Focused-ion-beam nanofabrication of near-infrared magnetic metamaterials. *Adv. Mater.* **17**, 2547-2549 (2005).

13. Rockstuhl, C., Zentgraf, T., Guo, H., Liu, N., Etrich, C., Loa, I., Syassen, K., Kuhl, J., Lederer, F. & Giessen, H. Resonances of split-ring resonator metamaterials in the near infrared. *Appl. Phys. B* **84**, 219–227 (2006).

14. Dolling, G., Enrich, C., Wegener, M., Zhou, J. F., Soukoulis, C. M. & Linden, S., Cut-wire pairs and plate pairs as magnetic atoms for optical metamaterials. *Opt. Lett.* **30**, 3198-3200 (2005)

15. Shalaev, V. M., Cai, W., Chettiar, U. K., Yuan, H.-K., Sarychev, A. K., Drachev, V. P. & Kildishev, A. V. Negative index of refraction in optical metamaterials. *Opt. Lett.* **30,** 3356-3358 (2005).

16. Zhang, S., Fan, W., Panoiu, N. C., Malloy, K. J., Osgood, R. M.& Brueck, S. R. J. Experimental Demonstration of Near-Infrared Negative-Index Metamaterials. *Phys. Rev. Lett.* **95**, 137404 (2005).

17. Dolling, G., Enrich, C., Wegener, M., Soukoulis, C. M. & Linden, S. Simultaneous Negative Phase and Group Velocity of Light in a Metamaterial. *Science* **312**, 892-894 (2006).

18. Shafiei, F., Monticone, F., Le, K. Q., Liu, X.-X., Hartsfield, T., Alu, A. & Li X. A subwavelength plasmonic metamolecule exhibiting magnetic-based optical Fano resonance. *Nature Nanotech.* **8**, 95–99 (2013).

19. Sheikholeslami, S. N., Alaeian, H., Koh, A. L. & Dionne, J. A. A Metafluid Exhibiting Strong Optical Magnetism. *Nano Lett.* (2013), DOI: 10.1021/nl401642z

20. Dolling, G., Wegener, M., Soukoulis, C. M. & Linden, S. Negative-index metamaterial at 780 nm wavelength. *Opt. Lett.* **32,** 53-55 (2007).

21. Cai, W., Chettiar, U. K., Yuan, H.-K., de Silva, V. C., Kildishev, A. V., Drachev, V. P. & Shalaev, V. M. Metamagnetics with rainbow colors. *Opt. Express* **15**, 3333-3341 (2007)

22. Xiao, S., Chettiar, U. K., Kildishev, A. V., Drachev, V. P. & Shalaev, V. M. Yellow-light negative-index metamaterials. *Opt. Lett.* **34**, 3478 (2009).

23. Lahiri, B., McMeekin, S. G., Khokhar, A. Z., De La Rue R. M. & Johnson, N. P. Magnetic response of split ring resonators (SRRs) at visible frequencies. *Opt. Express* **18**, 3210 (2010).

24. Garcıa-Meca, C., Hurtado, J., Marti, J., Martinez, A., Dickson, W. & Zayats, A. V. Low-loss multilayered metamaterial exhibiting a negative index of refraction at visible wavelengths. *Phys. Rev. Lett.* **106**, 067402 (2011).

25. Tomioka, T., Kubo, S., Nakagawa, M., Hoga, M. & Tanaka, T. Split-ring resonators interacting with a magnetic field at visible frequencies. *Appl. Phys. Lett.* **103**, 071104 (2013).

26. Evlyukhin, A. B., Reinhardt, C., Seidel, A., Luk'yanchuk, B. S. & Chichkov, B. N. Optical response features of Si-nanoparticle arrays. *Phys. Rev. B* **82**, 045404 (2010).

27. Garcia-Etxarri, A., Gomez-Medina, R., Froufe-Perez, L. S., Lopez, C., Chantada, L., Scheffold, F., Aizpurua, J., Nieto-Vesperinas, M. & Saenz, J. J. Strong magnetic response of submicron silicon particles in the infrared. *Opt. Express* **19**, 4815–4826 (2011).

28. Kuznetsov, A. I., Miroshnichenko, A. E., Fu, Y. H., Zhang, J. B. & Luk'yanchuk, B. Magnetic Light. *Sci. Rep.* **2**, 492 (2012).

29. Evlyukhin, A. B., Novikov, S. M., Zywietz, U., Eriksen, R. L., Reinhardt, C., Bozhevolnyi, S. I. & Chichkov, B. N. Demonstration of Magnetic Dipole Resonances of Dielectric Nanospheres in the Visible Region. *Nano Lett.* **12**, 3749–3755 (2012).

30. Geffrin, J. M., García-Cámara, B., Gómez-Medina, R., Albella, P., Froufe-Pérez, L. S., Eyraud, C., Litman, A., Vaillon, R., González, F., Nieto-Vesperinas, M., Sáenz, J. J., Moreno, F. Magnetic and electric coherence in forward- and back-scattered electromagnetic waves by a single dielectric subwavelength sphere. *Nat. Commun.* **3**, 1171 (2012).

31. Fu, Y. H., Kuznetsov, A. I., Miroshnichenko, A. E., Yu, Y. F., Luk'yanchuk, B. Directional visible light scattering by silicon nanoparticles. *Nat. Commun.* **4**, 1527 (2013).

32. Person, S., Jain, M., Lapin, Z., Sáenz, J. J., Wicks, G. & Novotny, L. Demonstration of zero optical backscattering from single nanoparticles. *Nano Lett.* **13**, 1367-1868 (2013).

33. Chen, W. T., Chen, C. J., Wu, P. C., Sun, S., Zhou, L., Guo, G.-Y., Hsiao, C. T., Yang, K.-Y., Zheludev, N. I. & Tsai, D. P. Optical magnetic response in three-dimensional metamaterial of upright plasmonic meta-molecules. *Opt. Express* **19**, 12837-12842 (2011).

34. Muehlig, S., Menzel, C., Rockstuhl, C. & Lederer, F. Multipole analysis of meta-atoms. *Metamaterials* **5**, 64–73 (2011).

35. Jackson, J. D. Classical Electrodynamics. Wiley, 3rd edition (1998).

36. Grahn, P., Shevchenko, A. & Kaivola, M. Electromagnetic multipole theory for optical nanomaterials. *New J. Phys.* **14**, 093033 (2012).

37. Kuznetsov, A. I., Evlyukhin, A. B., Gonçalves, M. R., Reinhardt, C., Koroleva, A., Arnedillo, M. L., Kiyan, R., Marti, O. & Chichkov, B. N. Laser fabrication of large-scale nanoparticle arrays for sensing applications. *ACS Nano* **5**, 4843–4849 (2011).

38. Leen, J. B., Hansen, P., Cheng, Y.-T. & Hesselink L. Improved focused ion beam fabrication of near-field apertures using a silicon nitride membrane. *Opt. Lett.* **33**, 2827-2829 (2008).

39. Gervinskas, G., Seniutinas, G., Rosa, L. & Juodkazis, S. Arrays of arbitrarily shaped nanoparticles: overlay-errorless direct ion write, *Adv. Optical Mater.* **1**, 456–459 (2013).

40. Ananth, M., Stern, L., Ferranti, D., Huynh, C., Notte, J., Scipioni, L., Sanford, C. & Thompson, B. Creating Nanohole Arrays with the Helium Ion Microscope. *Proc. SPIE* **8036**, 80360M (2011).

41. Scipioni, L., Ferranti, D. C., Smentkowski, V. S. & Potyrailo, R. A. Fabrication and initial characterization of ultrahigh aspect ratio vias in gold using the helium ion microscope, *J. Vac. Sci. Tech. B* **28**, C6P18 (2010).

42. Scholder, O., Jefimovs, K., Shorubalko, I., Hafner, C., Sennhauser, U. & Bona, G.-L. Helium focused ion beam fabricated plasmonic antennas with sub-5 nm gaps. *Nanotechnology* **24**, 395301 (2013).

43. Pickard, D.S. A distributed axis electron beam system for high-speed lithography and defect inspection. *PhD Thesis*, Stanford University (2006).

44. Luk'yanchuk, B. S., Miroshnichenko, A. E. & Kivshar, Yu. S. Fano resonances and topological optics: an interplay of far- and near-field interference phenomena**,** *J. Opt.* **15**, 073001 (2013).

45. Rahmani, M., Miroshnichenko, A. E., Lei, D. Y., Luk'yanchuk, B. S., Tribelsky, M. I., Kuznetsov, A. I., Kivshar, Yu. S., Francescato, Y., Giannini, V., Hong, M. & Maier, S.

A. Beyond the hybridization effects in plasmonic nanoclusters: diffraction-induced enhanced absorption and scattering, Small (2013), DOI: 10.1002/smll.201301419

46. Luk'yanchuk, B., Zheludev, N. I., Maier, S. A., Halas, N. J., Nordlander, P., Giessen, H. & Chong, C. T. The Fano resonance in plasmonic nanostructures and metamaterials. *Nature Mater.* **9,** 707-715 (2010).

47. Miroshnichenko, A. E., Flach, S. & Kivshar, Y. S. Fano resonances in nanoscale structures. *Rev. Mod. Phys.* **82,** 2257-2298 (2010).

48. Yang, J., Rahmani, M., Teng, J. H. & Hong, M. H. Magnetic-electric interference in metal-dielectric-metal oligomers: generation of magneto-electric Fano resonance. *Opt. Mater. Express* **2**, 1407-1415 (2012).

49. Kuznetsov, A. I., Koch, J. & Chichkov, B. N. Laser-induced backward transfer of gold nanodroplets. *Opt. Express* **17**, 18820–18825 (2009).

50. Kuznetsov, A. I., Evlyukhin, A. B., Reinhardt, C., Seidel, A., Kiyan, R., Cheng, W., Ovsianikov, A. & Chichkov, B. N. Laser-induced transfer of metallic nanodroplets for plasmonics and metamaterial applications. *J. Opt. Soc. Am. B* **26**, B130–B138 (2009).

51. Kuznetsov, A. I., Kiyan, R. & Chichkov, B. N. Laser fabrication of 2D and 3D microstructures from metal nanoparticles. *Opt. Express* **18**, 21198–21203 (2010).

**Acknowledgment**

Yiguo Chen (DSI) is acknowledged for help with laser nanoparticle fabrication at the initial stage of this work. This work was supported by the Singapore National Research Foundation project NRF2011NRF-CRP002-050, by the National University of Singapore funded project R263000674133, by the Agency for Science, Technology and Research (A*STAR) of Singapore: SERC Grant 102 152 0011, and by the Australian Research Council through the AEM Future Fellowship project FT110100037.

**Author contributions**

AIK contributed to initial idea generation, laser processing, CST simulations, and spectral characterization, he coordinated the whole work and wrote the paper; AEM contributed to initial idea generation, performed CST simulations and multipole decomposition, and contributed to the manuscript preparation; YHF performed spectral characterization; VigVis performed HIM milling and imaging; MR performed e-beam nanofabrication; VytVal and ZYP performed laser processing; YK supervised numerical simulations; DSP supervised HIM work; BL supervised the whole work. All authors read and corrected the manuscript before the submission.

**Additional information**

Competing financial interests: The authors declare no competing financial interests.

**Figure legends**

**FIGURE 1.** Schematic representation of the split-ball resonator and geometry of spectral measurements.

**FIGURE 2.** Numerical simulations of optical properties of SBRs. **(a)** Total scattering efficiencies of the SBR formed by a gold sphere with diameter of 200 nm and a cut with 50 nm width and 100nm depth. The SBR is excited by a plane wave directed from top antiparallel to Z axis (solid curves) or from the side facing the cut along Y axis (dashed curves). Polarization of the incident light is perpendicular or parallel to the cut. Black dotted curve represents scattering efficiency of the same particle without a cut. Grey curve is absorption efficiency of the SBR for top excitation with polarization perpendicular to the cut. Inset is the schematic illustration of the SBR and directions of the incident plane waves. **(b)-(e)** Distributions of electric **(b)&(d)** and magnetic **(c)&(e)** field around the SBR at electric **(b)&(c)** (622 nm, polarization is parallel to the cut) and *LC* **(d)&(e)** (882 nm, polarization is perpendicular to the cut) resonance wavelengths. **(f)-(i)** Decomposition of the scattered field of SBR into multipole moments based on the vector spherical harmonics. SBR is excited from the top **(f)&(g)** or from the side facing the nanocut **(h)&(i)**. Incident light polarization is perpendicular **(f)&(h)** or parallel **(g)&(i)** to the cut. All the electric and magnetic spherical multipole harmonics of the first and the second order are calculated and plotted in the figure. Only harmonics which have a noticable contribution are marked by corresponding coefficients ($b_{10}$ – magnetic dipole, $a_{10}$ and $a_{11}$ – electric dipoles). $a_{11}$ in all plots stays for a sum of $a_{11}$ and $a_{1-1}$ mode contributions.

**FIGURE 3.** Dependence of scattering efficiency of the SBR on the cut depth at a fixed cut width of 20 nm **(a)** and on the cut width at a fixed cut depth of 100 nm **(b)**. The incident plane wave is directed from top of the SBR with light polarization perpendicular to the cut (see Fig.1). The gold sphere diameter is 200 nm.

**FIGURE 4.** Experimental realization of gold SBRs. **(a)&(b)** – HIM images of two gold nanoparticles with similar diameters of 170 nm and cuts width of 15 nm. Cut depth in **(a)** is larger than in **(b)**, which is obtained by 25% longer helium ion beam milling time. **(c)&(d)** Scattering spectra of the particles shown in **(a)&(b)** for incident light with s (perpendicular to the cut) and p (parallel to the cut) polarizations.

**FIGURE 5.** Experimental realization of silver SBRs. **(a)-(c)** HIM images of three silver nanoparticles with diameters of 170 nm **(a)** and 80 nm **(b)&(c)** and a cut width of 15 nm **(a)** and 12 nm **(b)&(c)**. Particle in **(c)** is chemically synthetized. **(d)-(f)** Scattering spectra of the nanoparticles shown in **(a)-(c)** for incident light with s (perpendicular to the cut) and p (parallel to the cut) polarizations.

**FIGURE 6.** Poynting vector distribution around gold SBR with 200 nm diameter and a cut with 50 nm width and 100 nm depth (similar to Fig.2) at *LC* resonance wavelength of 882 nm **(a)-(c)** and scattering minimum wavelength of 710 nm **(d)-(f)**. **(a)&(d)** show XZ cross-section plane in the middle of the sphere. **(b)&(e)** show YZ cross-section plane in the middle of the sphere. **(c)&(f)** show separatrix absorption surfaces (limited by yellow separatrix lines) of the SBR in XY cross-section plane. Colour map represents distribution of the modulus of the Poynting vector around the nanoparticle. Red lines show separatrices of the energy flow [37, 38]. Green circles highlight the nanoparticle location.

**Fig. 1**

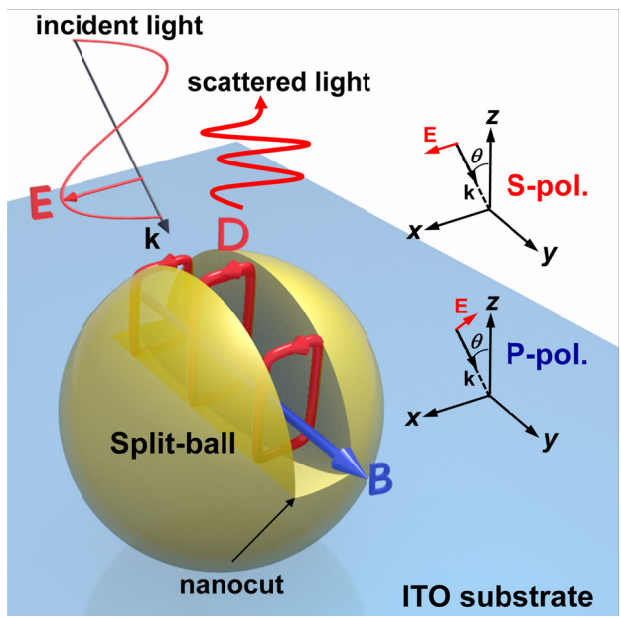

**Fig. 2**

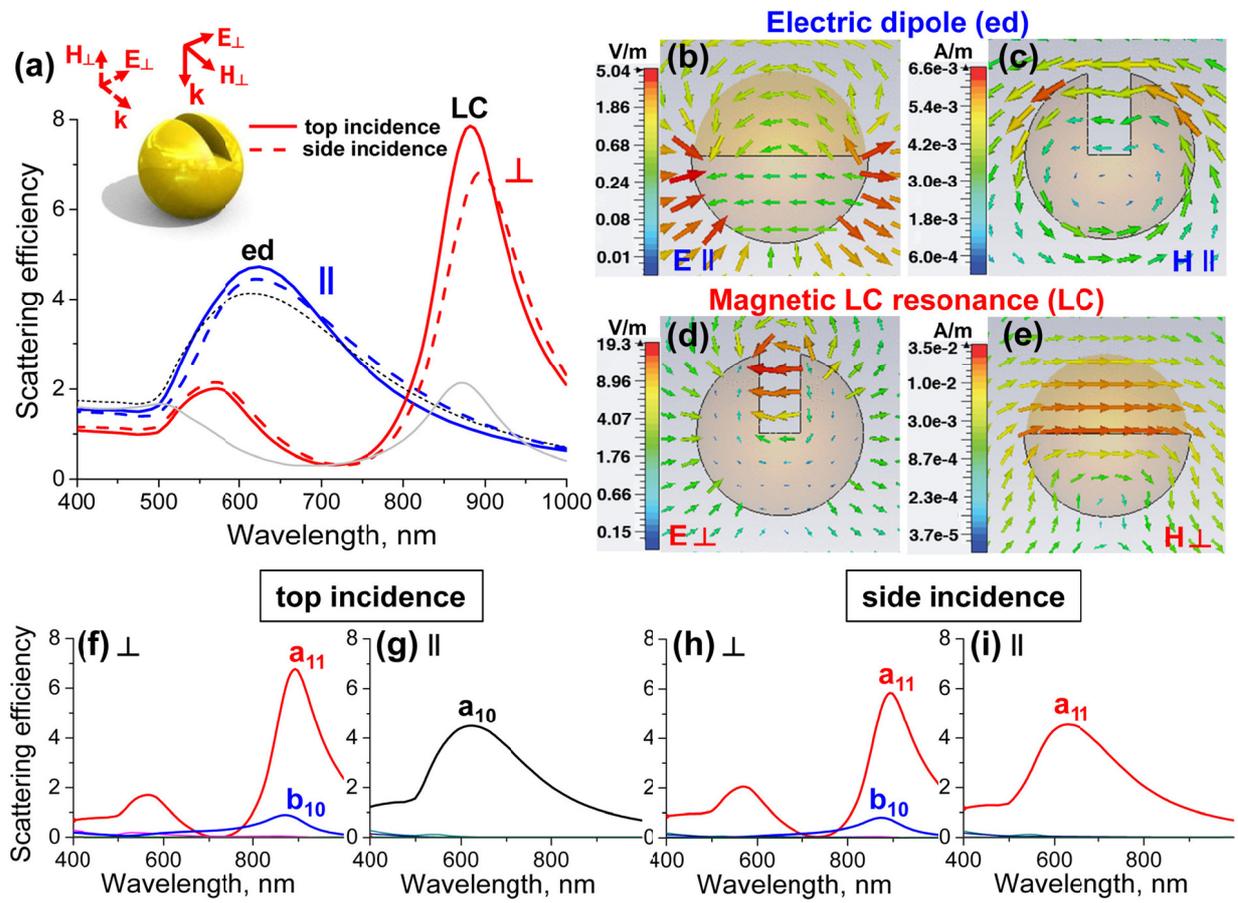

**Fig.3**

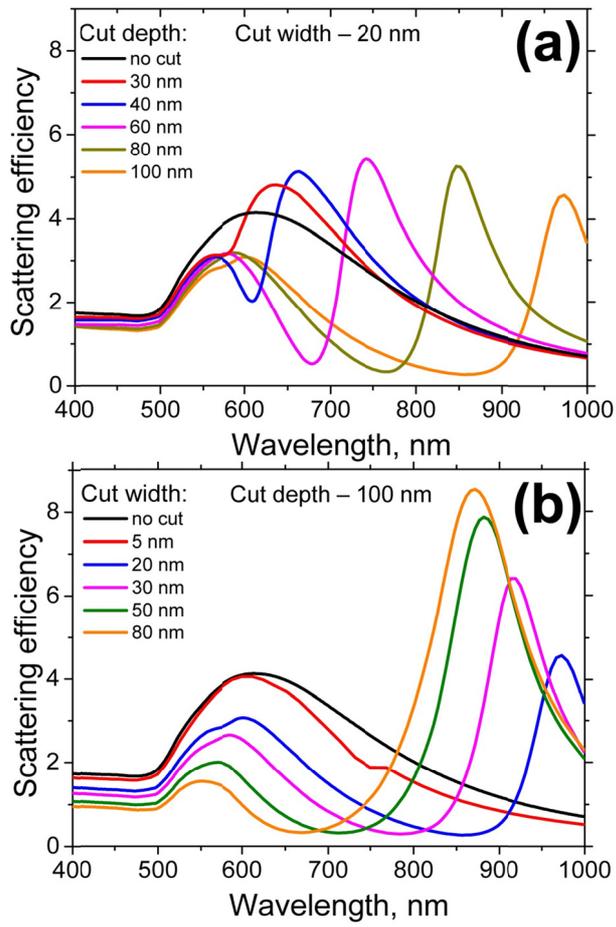

**Fig.4**

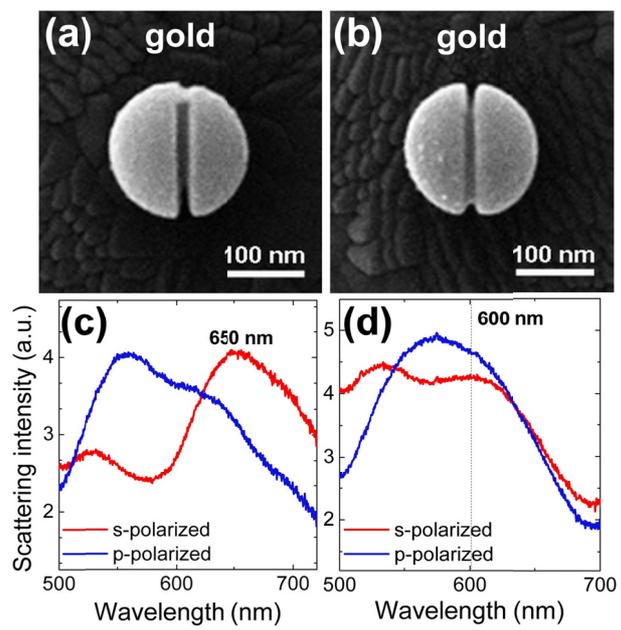

**Fig.5**

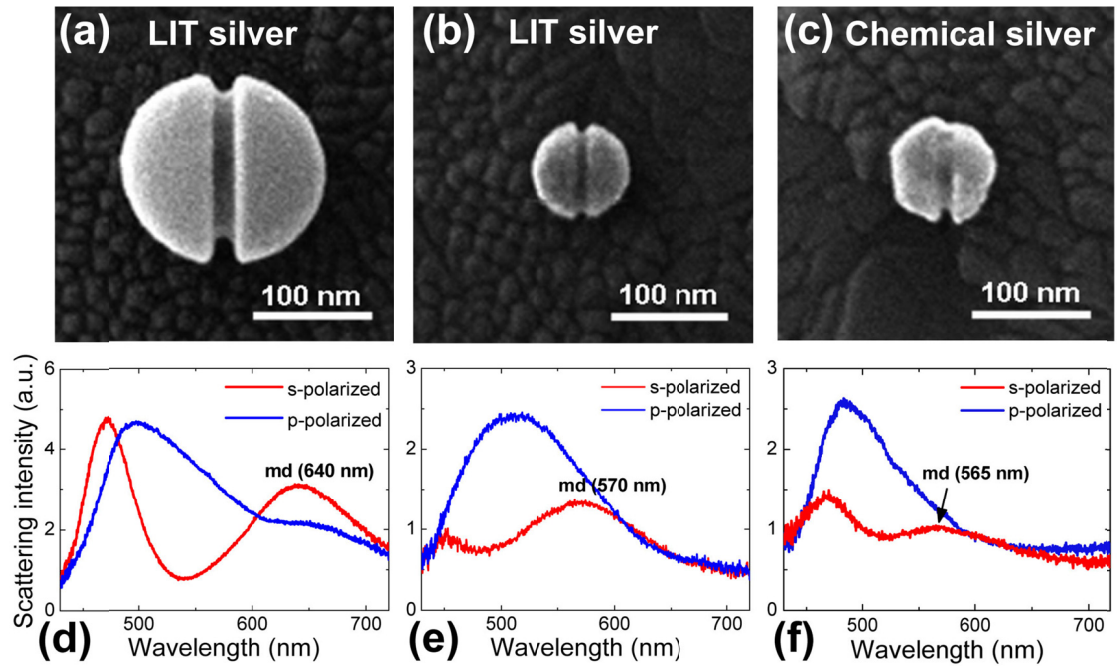

Fig.6

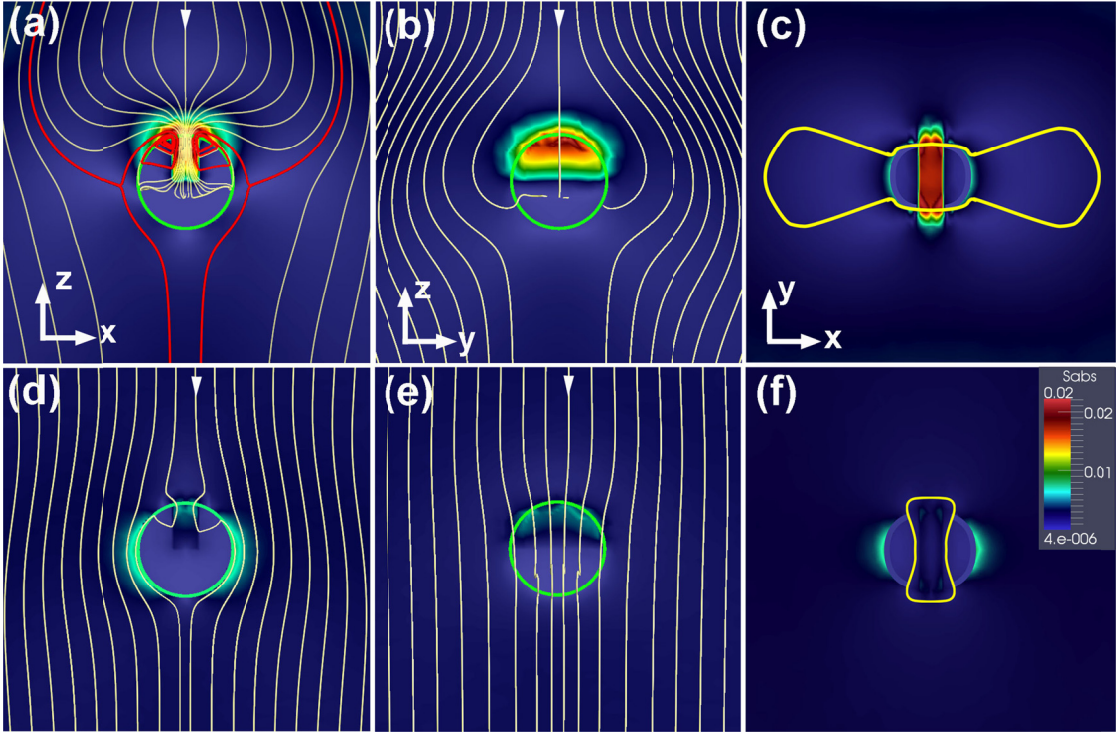

# Supplementary Information

# Split-ball resonator

*Arseniy I. Kuznetsov,[1,]\* Andrey E. Miroshnichenko,[2] Yuan Hsing Fu,[1] Vignesh Viswanathan,[3] Mohsen Rahmani,[1] Vytautas Valuckas,[1,3] Zhen Ying Pan,[1] Yuri Kivshar,[2] Daniel S. Pickard,[3] and Boris Luk'yanchuk,[1]*

1. Data Storage Institute, A*STAR (Agency for Science, Technology and Research), 5 Engineering Drive 1, 117608, Singapore
2. Nonlinear Physics Centre, Research School, of Physics and Engineering, Australian National University, Canberra ACT 0200, Australia;
3. Department of Electrical and Computer Engineering, National University of Singapore, 1 Engineering Drive 2, 117576, Singapore

*Corresponding author:  arseniy_k@dsi.a-star.edu.sg

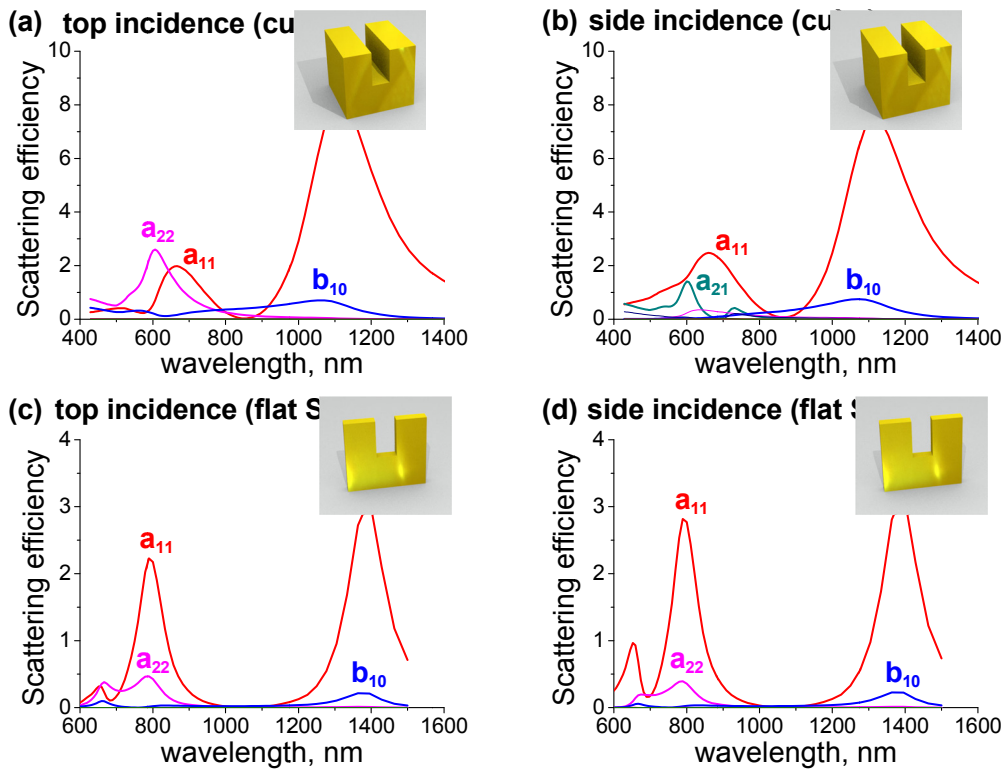

**Supplementary Figure S1:** Partial light scattering efficiencies by different spherical multipole modes excited inside different split-ring resonators. **(a)&(b)** Scattering by a 3D gold cube with a cut on the top. The cube dimentions are 200×200×200 nm$^3$, the cut has a width of 50 nm and depth of 100 nm. The incoming plane wave is directed from the top **(a)** and from the side facing

the cut **(b)**. **(c)&(d)** Scattering by a flat gold split-ring standing upright. The SRR has dimensions of 200×200×20 nm$^3$, the cut has a width of 50 nm and depth of 100 nm. The incoming plane wave is directed from the top **(c)** and from the side facing the cut **(d)**. Indexes $a_{lm}$ and $b_{lm}$ correspond to electric and magnetic spherical multipoles of different order, respectively. In particular, $a_{11}$ corresponds to electric dipole mode, $b_{10}$ to magnetic dipole mode, $a_{22}$ and $a_{21}$ are electric quadrupole modes. In case $m$ is not zero, $a_{lm}$ in the plots stays for a sum of $a_{lm}$ and $a_{l-m}$ mode contributions. The ratio of the maximum scattering efficiency by magnetic dipole mode $b_{10}$ to the maximum scattering efficiency by electric dipole mode $a_{11}$ is: 0.084 in (a); 0.099 in (b); 0.069 in (c); and 0.071 in (d). These numbers are lower than the corresponding values for split-ball resonator with similar gap size: 0.132 in Fig.2f and 0.138 in Fig.2h. All the electric and magnetic spherical multipole harmonics of the first and the second order are calculated and plotted in the figures. Only harmonics which have a noticable contribution are marked.